\documentclass[aps,pre,twocolumn,showpacs]{revtex4}
\usepackage{graphicx}
\begin{document}
\newcommand{\ud}{{\mathrm d}}
\newcommand{\sech}{\mathrm{sech}}

\title{Very large stochastic resonance gains in finite sets of interacting identical subsystems driven by subthreshold rectangular pulses}


\author{David Cubero}
\author{Jes\'us Casado-Pascual} \author{Jos{\'e}
G{\'o}mez-Ord{\'o}{\~n}ez} \author{Jos{\'e} Manuel Casado}
\author{Manuel Morillo} \affiliation{F\'{\i}sica Te\'orica, Universidad
de Sevilla, Apartado de Correos 1065, Sevilla 41080, Spain}


\date{\today}

\begin{abstract}
We study the phenomenon of nonlinear stochastic resonance (SR) in a
complex noisy system formed by a finite number of interacting subunits
driven by rectangular pulsed time periodic forces. We find that very
large SR gains are obtained for subthreshold driving forces with
frequencies much larger than the values observed in simpler
one-dimensional systems. These effects are explained using simple
considerations.
\end{abstract}


\maketitle

The phenomenon of stochastic resonance (SR) seems to be important in
a wide variety of contexts in physics, chemistry, and the life
sciences \cite{gamjun98}.  A lot of work has been devoted to the
study of SR both in simple \cite{han02} and complex systems
\cite{junmay95}, such as in ion channel assemblies
\cite{schigoy01} or globally coupled networks of noisy neural
elements \cite{zhokur03}, to name a few examples. In this
work, we consider a complex system formed by a finite number of $N$
coupled noisy bistable subsystems, the attention to finite sets
being inspired by the fact that certain processes in neuroscience
seem to involve a rather small number of subsystems \cite{abarab96}.

The signal-to-noise ratio (SNR) and the SR gain are two common
quantifiers used to characterize the SR response of noisy systems
driven by time-periodic forces. We will study SR effects on a
collective variable of finite sets of interacting subunits driven by
periodic rectangular pulses. Our results show a tremendous
enhancement of SR effects in this collective variable with respect to
those observed in single unit systems.

Let us consider a set of $N$ interacting subsystems, each one of them
characterized by a single degree of freedom $x_i$ ($i=1,\ldots,N$),
whose dynamics is governed by the Langevin equations
\cite{komshi75,deszwa78,casgom06}
\begin{equation}
\dot{x}_i=x_i-x_i^3+\frac{\theta}{N}\sum_{j=1}^N(x_j-x_i)+\xi_i(t)+F(t),
\label{eq:lang}
\end{equation}
where $\xi_i(t)$ are Gaussian white noises with zero average and
$\langle \xi_i(t)\xi_j(s)\rangle=2D\delta_{ij}\delta(t-s)$, $\theta$ is
the parameter defining the strength of the interaction between
subsystems, and $F(t)$ is an external driving force of period $T$. In
this work, we will restrict ourselves to forces of the type
\cite{casgom04}
\begin{equation}
F(t)=\left\{ \begin{array}{lll}
A &;& 0\le t<t_c, \\
0 &;& t_c\le t < T/2, \\
-A &;& T/2 \le t < T/2+t_c, \\
0 &;& T/2+t_c \le t < T.
            \end{array} \right.
\label{eq:force:def}
\end{equation}
The parameter $r=2t_c/T$, usually called {\em duty cycle}, measures
the fraction of a period during which this driving force has a
nonvanishing value.  In addition, we will only consider {\em
subthreshold} amplitudes $A$, so that the driving force
(\ref{eq:force:def}) cannot induce sustained oscillations between
dynamical attractors in the absence of noise. The model described by
Eq.~(\ref{eq:lang}) (without the external periodic driving) was used
years ago by Kometani and Shimizu \cite{komshi75} as an empirical
model to describe muscle contraction. Later on, Desai and Zwanzig
\cite{deszwa78} gave a more detailed statistical mechanical
description in the asymptotic $N \rightarrow \infty$ limit and used
it to model order-disorder transitions. The addition of an external
driving can, in principle, be used to describe the phenomenology of
a forced contracting muscle.

We focus on the collective variable  $S(t)$ defined as
\begin{equation}
S(t)=\frac{1}{N}\sum_{i=1}^N x_i(t),
\label{eq:def:S}
\end{equation}
which  has previously been used \cite{casgom06} in the global
analysis of coupled bistable systems. It might be considered as the
total output process of a parallel array of N indentical interacting
subunits, subject to independent noise sources, $\xi_i(t)$ and the
same external forcing $F(t)$.
Its signal-to-noise ratio ($R_\mathrm{out}$) is defined in the
usual way as,
\begin{equation}
 R_\mathrm{out}=\frac{Q_u}{Q_l},
\end{equation}
with
\begin{eqnarray}
Q_u&=&\frac{2}{T}\int_0^T d\tau\,
C_{\mathrm{coh}}(\tau)\cos(\Omega\tau) \\
\label{ql}
Q_l&=&\frac{2}{\pi}\int_0^\infty d\tau\,
C_{\mathrm{incoh}}(\tau)\cos(\Omega\tau).
\end{eqnarray}
where $\Omega=2\pi/T$, $C_{\mathrm{coh}}(\tau)=\frac{1}{T}\int_0^T
dt\langle S(t+\tau)\rangle_{\infty}\langle S(t)\rangle_{\infty}$ and
$C_{\mathrm{incoh}}(\tau)=C(\tau)-C_{\mathrm{coh}}(\tau)$ with
$C(\tau)=\frac{1}{T}\int_0^T dt\langle
S(t+\tau)S(t)\rangle_{\infty}$. 

For a set of $N$ coupled linear oscillators driven by the external
driving force $F(t)$ and subject to the noise terms $\xi_i(t)$ as in
Eq. (\ref{eq:lang}), the SNR of the corresponding collective process, $R_\mathrm{out}^{(L)}$, coincides with that of the random process formed
by the arithmetic mean of the individual noise terms $\xi_i(t)$ plus the
deterministic driving force $F(t)$, namely, $F(t)+\xi(t)$ with $\xi(t)=N^{-1}\sum_{i=1}^N\xi_i(t)$.  The process $\xi(t)$ is a Gaussian white noise of effective strength $D/N$. Then, it is easy to prove that
\begin{equation}
R_\mathrm{out}^{(L)}=\frac{2 A^2 N [1-\cos(\pi r)]}{\pi D}.
\label{eq:Rin}
\end{equation}
Thus, for our nonlinear case, it seems convenient to
analyze the SR gain, $G$, defined as \cite{casgom06}
\begin{equation}
\label{gain} G=\frac{R_\mathrm{out}}{R_\mathrm{out}^{(L)}},
\end{equation}
which compares the SNR of a non-linear system with that of a linear system subject to the same stochastic and deterministic forces. 

We have carried out extensive simulations of the Langevin equations
(\ref{eq:lang}) with the external driving (\ref{eq:force:def}). In all
cases reported here the coupling strength is fixed to $\theta=0.5$ and
the subthreshold driving amplitude to $A=0.3$. There is nothing special
about this particular $\theta$ value. The qualitative results would be
the same for any other value of $\theta\ne 0$.

\begin{figure}
\includegraphics[width=7cm]{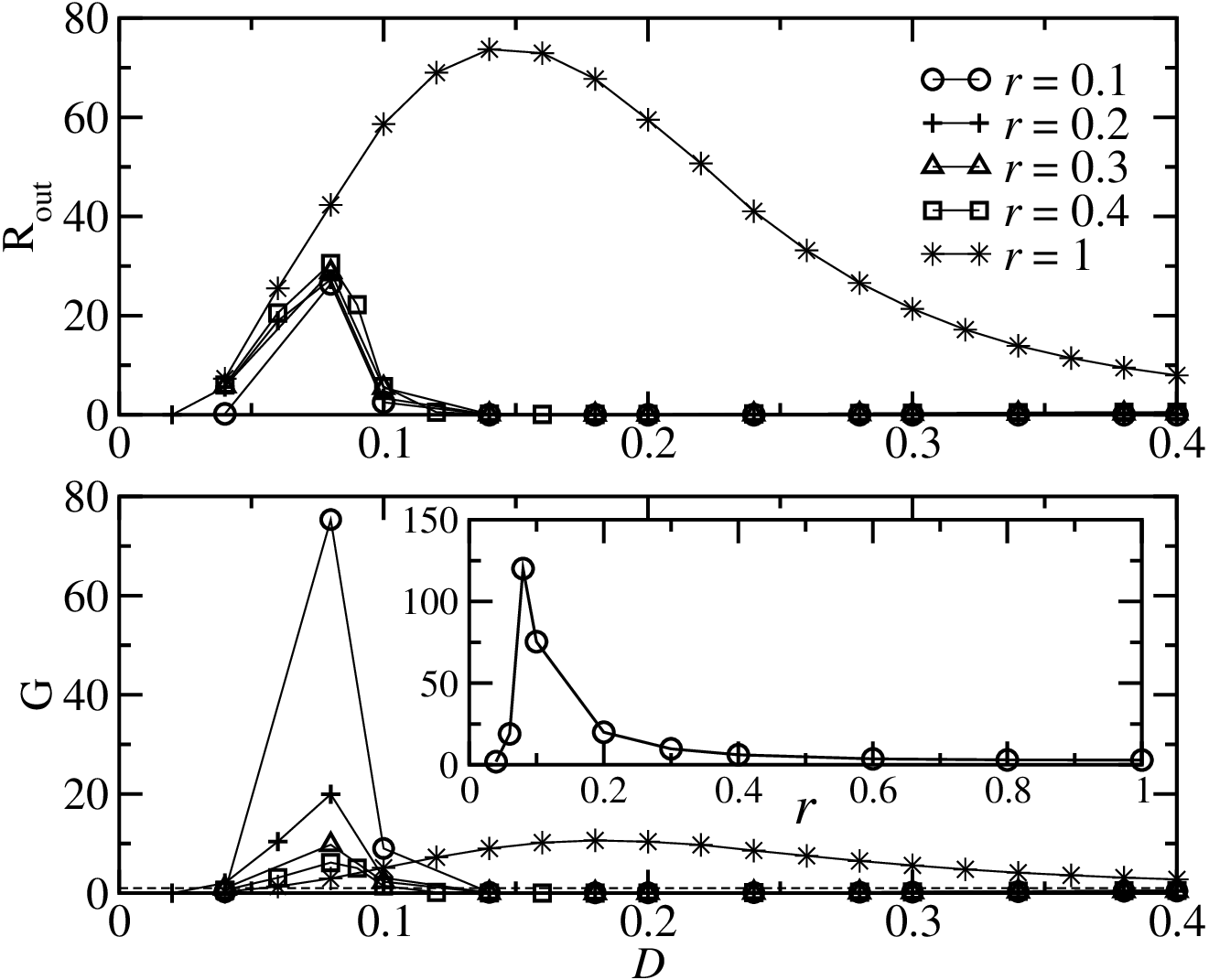}
\caption{
\label{fig:gain:N10}
Dependence of the signal-to-noise ratio $R_\mathrm{out}$ and the SR gain
 $G$ for a set of $N=10$ identical subsystems and an external driving of
 frequency $\Omega=0.01$ for several values of the duty cycle $r$: 0.1
 (open circles), 0.2 (crosses), 0.3 (triangles), 0.4 (squares), and 1
 (stars). Inset shows the SR gain as a function of $r$ for a fixed noise
 strength $D=0.08$. The lines have been drawn as a guide to the eye.  }
\end{figure}

In Fig.~\ref{fig:gain:N10} we show the collective signal-to-noise
ratio and SR gain as a function of the noise strength $D$ for a set
of $N=10$ identical subsystems, a driving fundamental frequency
$\Omega=0.01$ and several values of the duty cycle $r$. It can be
seen that, while the SNR curves are nearly identical for $r\le 0.4$,
the SR gain increases drastically, reaching a very large value for
$r=0.1$. This dramatic increase is easily understood by taking into
account the observed almost constant behavior of $R_\mathrm{out}$ and the fact that $R_\mathrm{out}^{(L)}$, Eq.~(\ref{eq:Rin}), decreases monotonically with
$r$. However, for sufficiently small $r$, $R_\mathrm{out}$ must decrease faster than $R_\mathrm{out}^{(L)}$, because the interval $t_c$ becomes smaller than
the time it takes for the system to react to a constant force of
amplitude $A$, and thus, the driving produces almost no effect in the
system. This behavior
is shown in the inset of Fig.~\ref{fig:gain:N10}, where it can be
seen that the SR gain decreases for $r \leq 0.08$. Also, as seen in
Fig.~\ref{fig:gain:N10}, when $r=1$ (rectangular driving signal),
$G$ reaches a peak at a noise value $ D\approx 0.2$. Even though
$R_\mathrm{out}^{(L)}$ is as large as it can be, the huge increase of
$R_\mathrm{out}$ is enough to overcome the increase in
$R_\mathrm{out}^{(L)}$, yielding a substantial value for the SR gain.

\begin{figure}
\includegraphics[width=7cm]{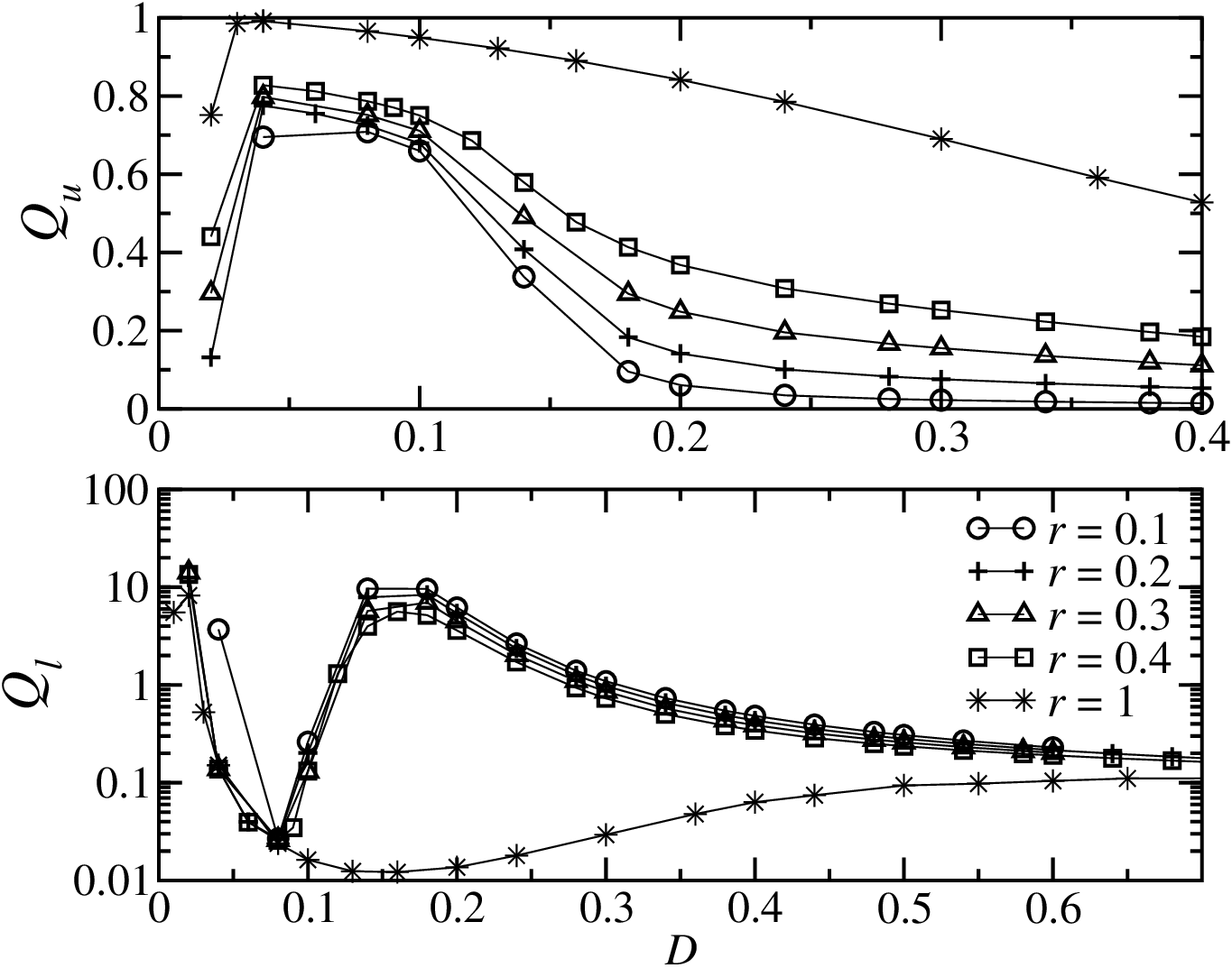}
\caption{
\label{fig:Qs}
Dependence of the denominator and the numerator of the collective
 signal-to-noise ratio for the same parameter values as in
 Fig~\ref{fig:gain:N10}. The lines have been drawn as a guide to the
 eye.  }
\end{figure}

The very large gain values in Fig.~\ref{fig:gain:N10} for pulses with
short duty cycles are observed only for a small range of noise strengths
around $D\approx0.08$. In Fig.~\ref{fig:Qs} we present the behavior of $Q_l$ and $Q_u$ with $D$. Notice that around  $D\approx0.08$ there is a strong reduction of two orders of magnitude in the level of fluctuations of the collective variable as measured by $Q_l$. Therefore, the large SR effects quantified by
the SNR and the SR gains are essentially due to the very large reduction
of the fluctuation spectrum of the output signal at the fundamental
driving frequency for a range of noise values.

\begin{figure}
\vspace*{1em}
\includegraphics[width=7cm]{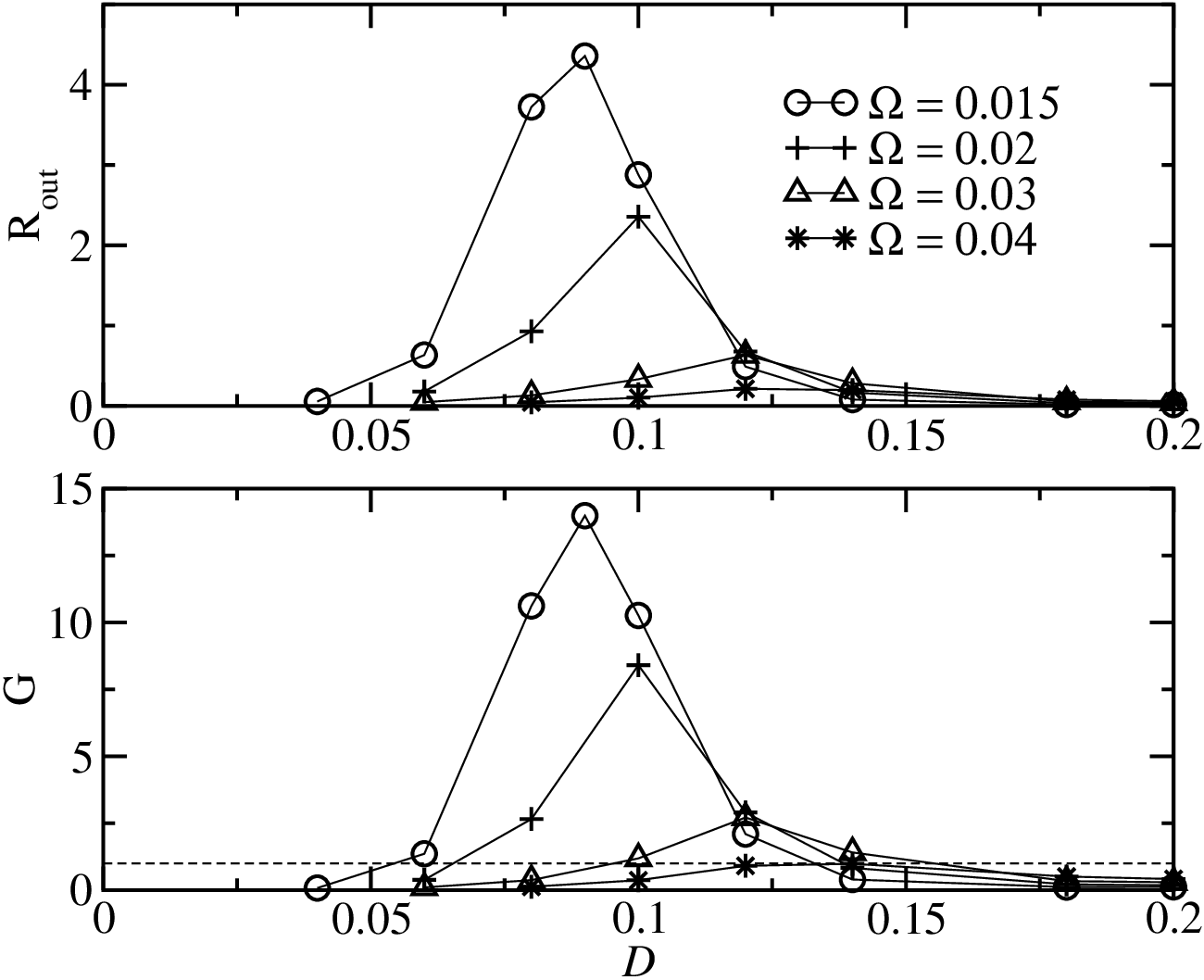}
\caption{
\label{fig:gain:N10:Om}
Dependence of the signal-to-noise ratio and the SR gain for a set of
 $N=10$ identical subsystems and duty cycle $r=0.1$ for several values
 of the driving frequency $\Omega$: 0.015 (open circles), 0.02
 (crosses), 0.03 (triangles), and 0.04 (squares). Solid lines have been
 drawn as a guide to the eye. The horizontal dashed line marks the unity
 for the SR gain.  }
\end{figure}

We have also analyzed the SNR and the SR gain for different values
of $\Omega$. Our results for $R_\mathrm{out}$ and $G$ as a function
of $D$ are depicted in Fig.~\ref{fig:gain:N10:Om} for a set of
$N=10$ interacting identical subunits driven by rectangular pulses
with duty cycle $r=0.1$ and several values of the fundamental
driving frequency $\Omega$: 0.015 (open circles), 0.02 (crosses),
0.03 (triangles), and 0.04 (squares). As we increase the driving
frequency $\Omega$, the SR gain is gradually reduced. For a
sufficiently large driving frequency (pulses of very short
duration), gains larger than unity are not observed for any value of
the noise strength.


As mentioned above, the large SNR values observed in
Fig.~\ref{fig:gain:N10} are related to the existence of a sharp minimum
of $Q_l$ at a certain noise value.  As observed in Fig.~\ref{fig:Qs},
this value is $D\approx 0.08$ for a driving signal with short duty
cycle, amplitude $A=0.3$ and fundamental frequency $\Omega =0.01$. For
other amplitude and frequency, the location of the minimum will change,
although the mechanisms leading to the existence of this minimum will be
qualitatively the same.

\begin{figure}
\includegraphics[width=7cm]{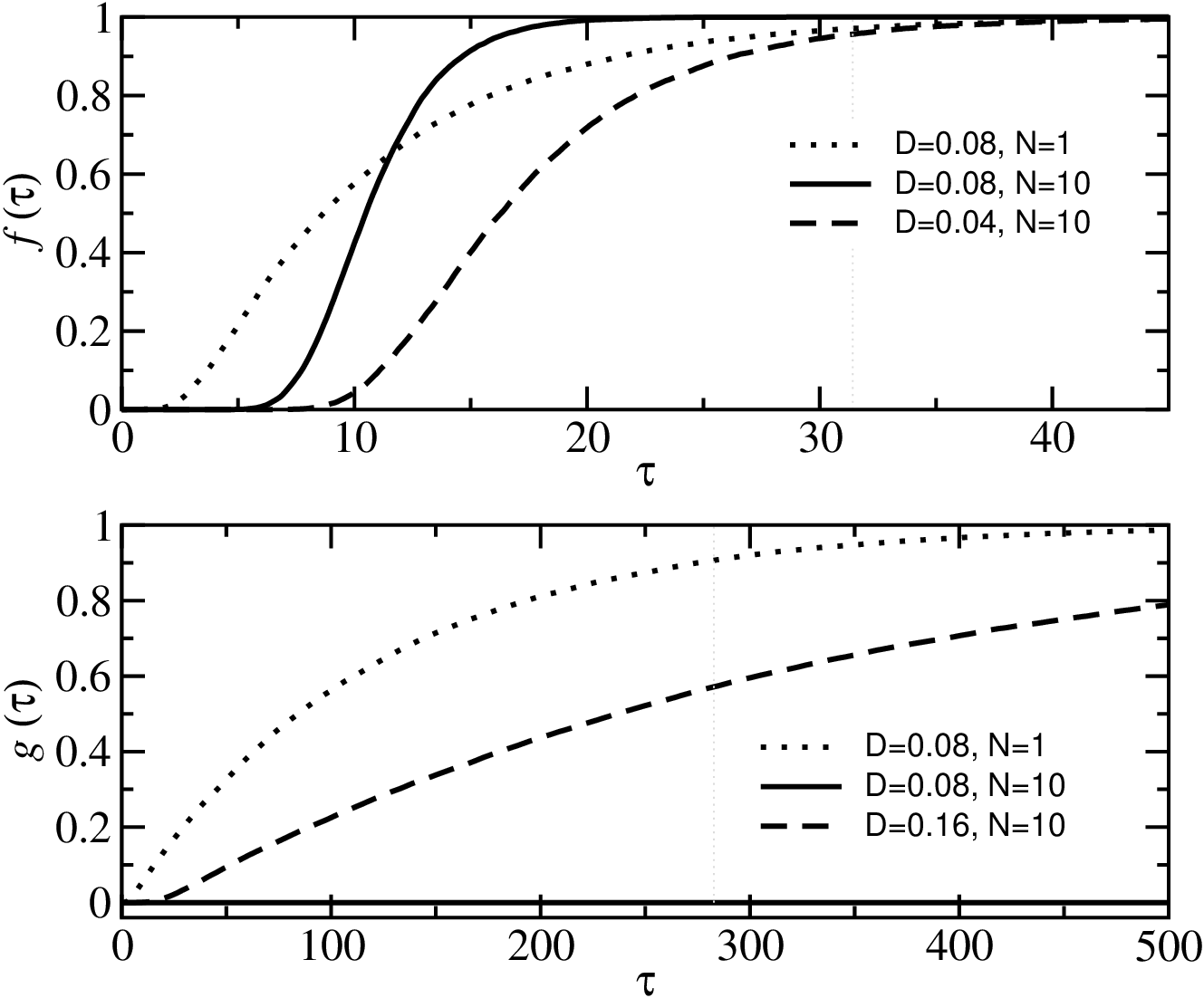}
\caption{
\label{fig:gain:tau}
(a) Probability that the collective variable $S(t)$ has jumped before a
 time $\tau$ to the attractor favored by an external constant driving of
 amplitude $A=0.3$ starting from the opposite attractor. The solid line
 corresponds to a system with $D=0.08$ and $N=10$, the dotted line to
 $D=0.08$ and $N=1$, and the dashed line to $D=0.04$ and $N=10$. (b)
 Probability that $S(t)$ jumps before a time $\tau$ to the opposite
 attractor in the absence of driving.  The solid line corresponds to a
 system with $D=0.08$ and $N=10$ (being almost zero for the range of
 times shown), the dotted line to $D=0.08$ and $N=1$, and the dashed
 line to $D=0.16$ and $N=10$.  }
\end{figure}

\begin{figure}
\vspace*{1em}
\includegraphics[width=7cm]{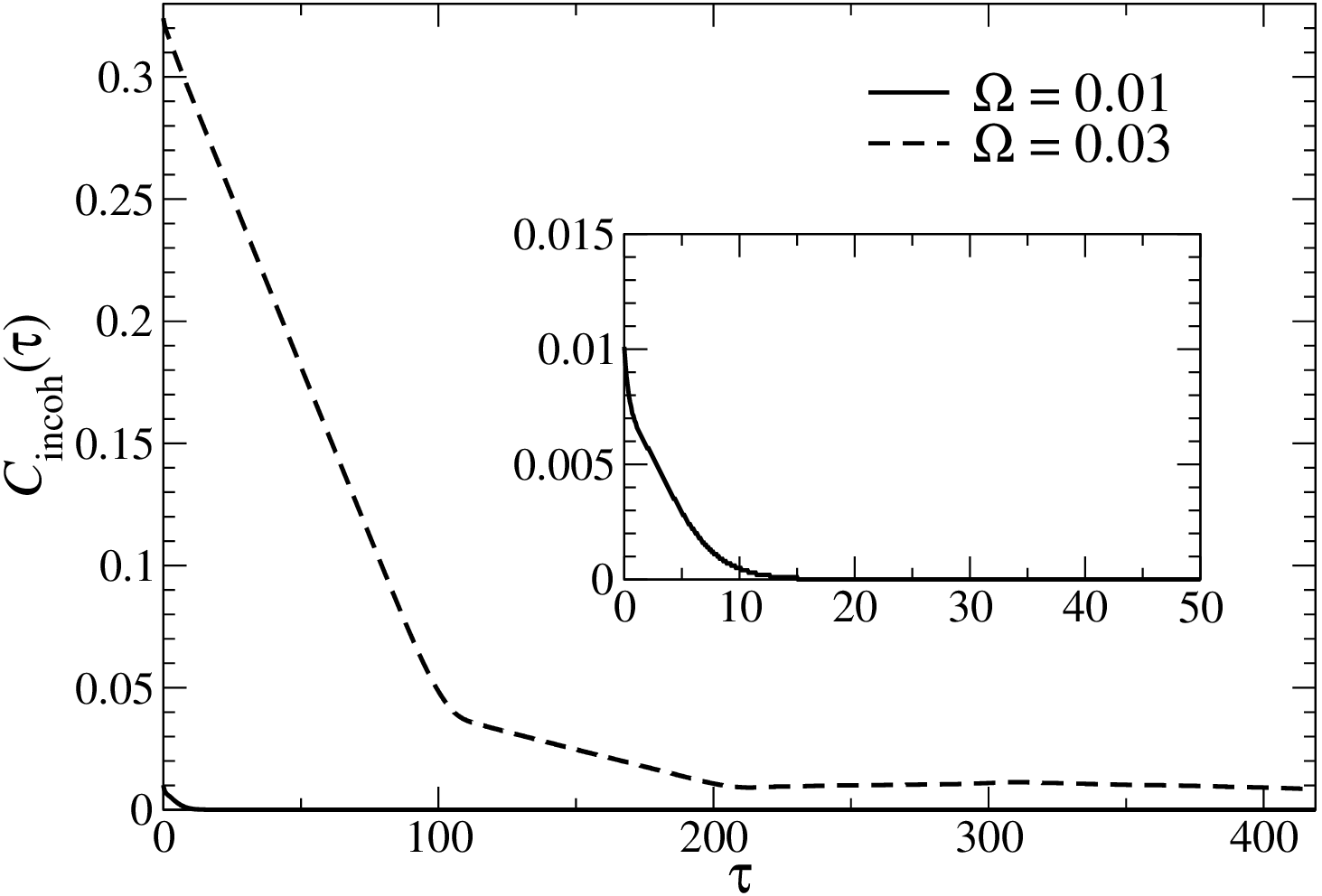}
\caption{
\label{fig:Cincoh}
Temporal behavior of the incoherent part of the correlation function of
 the collective variable for a system of $N=10$ identical subsystems
 with $D=0.08$ and $r=0.1$. The solid lines correspond to $\Omega=0.01$
 and the dashed line to $\Omega=0.03$. The inset is a magnification of
 the case $\Omega=0.01$ displayed for the sake of clarity.  }
\end{figure}

To understand the $Q_l$ dependence on $D$, it is important to analyze in
detail the dynamics imposed by the external driving force in
Eq. (\ref{eq:force:def}). For the $D$ and $\theta$ values of interest to
the present discussion, simulations show that when the driver is absent,
the collective variable $S(t)$ performs a noise induced random movement
between the two symmetric attractors of the dynamics, which we will
regard as located at $\pm S_0$. During a time interval of duration
$t_c$, the external force $F(t)=A$ favors the positive attractor. Thus,
if $S(t)$ was at the negative attractor at the beginning of the
interval, the external forcing will drive it to the positive one in a
random time that we will denote by $\Upsilon_1$. Certainly, different
realizations of the noise will yield different values of
$\Upsilon_1$. Running simulations with many independent trajectories, we
have computed the probability $\mathrm{Prob}(\Upsilon_1<\tau)=f(\tau)$
that the variable $S(t)$ has jumped before a time $\tau$ to the
attractor favored by a constant driving of amplitude $A$. In
Fig. \ref{fig:gain:tau}~(a) we present $f(\tau)$ for the set of
parameters relevant to the discussion: $A=0.3$, $D=0.08$, and $N=10$
(solid line). It can be seen that for the case $r=0.1$ and
$\Omega=0.01$, (thus, $t_c\approx31.4$), a transition between the
attractors for $\tau=t_c$ is performed with probability almost
unity. Then, during the rest of the half-period, an interval of duration
$T/2-t_c$, the external force is zero and the system is free to jump
between the attractors due solely to noise. Let us now denote by
$\Upsilon_2$ the random time it takes to jump from one attractor to the
opposite one when $F(t)=0$. Figure \ref{fig:gain:tau}~(b) shows the
probability $\mathrm{Prob}(\Upsilon_2<\tau)=g(\tau)$ that this jump has
taken place before a time $\tau$ vs. $\tau$. Since for $r=0.1$ and
$\Omega=0.01$ we have $T/2-t_c\approx282.7$, it can be checked in
Fig. \ref{fig:gain:tau}~(b) that almost no transitions take place during
this time interval under these conditions.  We can carry out the same
analysis for the symmetric situation during the second half-period of
the driving force. Consequently, $S(t)$ performs a neat trajectory
between its attractors with transitions induced systematically every
half-period by the external driving when it has a nonvanishing value.
As a result, we would expect $C_\mathrm{incoh}(0)$, itself an average of
the second cumulant of $S(t)$ over a period, to be of the order of the
effective noise $D/N$, and $C_\mathrm{incoh}(\tau)$ short-lived. This
is, in fact, what is observed in Fig.~\ref{fig:Cincoh} (see inset).

The curves in Fig. \ref{fig:gain:tau}(a) indicate that the probability
of transitions between the attractors before $t_c$ out of the most
unstable attractor is smaller for $D=0.04$ than for $D=0.08$. This is to
be expected as the probability of transitions out of an attractor goes
to zero when $D\to0$ with subthreshold driving amplitudes. As a result,
$C_\mathrm{incoh}(0)$ and the decay time of $C_\mathrm{incoh}(\tau)$
increase as $D$ gets smaller than $D=0.08$.  Consequently, noticing its
definition in Eq. (\ref{ql}), $Q_l$ has to increase as $D$ is reduced
from $D=0.08$.

Fig.~\ref{fig:Qs} shows that for noise values larger than $D\approx0.08$ the $Q_l$ dependence on $D$
for inputs with short duty cycles is rather different from the one
observed for a rectangular input ($r=1$). This fact indicates that for short duty cycles and these larger noise values, the time intervals during which $F(t)=0$ are crucial for the system response. By comparing the $g(\tau)$ curves for $D=0.08$ and $D=0.16$ in Fig. \ref{fig:gain:tau}~(b), we see that the probability of transitions between attractors during the time interval $T/2-t_c$ is negligible, whereas there is a non-negligible probability for $D=0.16$. Therefore, the bimodal character of the probability
distribution for $S(t)$ is enhanced as $D$ increases from $D\approx0.08$
to $D\approx0.16$ and one can then conclude that $C_\mathrm{incoh}(0)$
and the decay time of $C_\mathrm{incoh}(\tau)$ increase as $D$
increases. Consequently, for short duty cycles, $Q_l$ also increases as
$D$ increases within the interval $D\approx0.08$ to $D\approx0.16$. On
the other hand, for $r=1$ and this same range of noise values, the
situation is different because the driving signal never vanishes. Then,
transitions out of the most unstable attractor happen more frequently as
noise increases [$f(\tau)$ saturates at $1$ earlier], and this leads to
a decrease of both $C_\mathrm{incoh}(0)$ and the decay time of
$C_\mathrm{incoh}(\tau)$ as $D$ increases, with the corresponding
decrease in $Q_l$.

Similarly, raising the driving fundamental frequency has a drastic effect on the behavior of $C_\mathrm{incoh}(\tau)$ with respect to its optimal behavior discussed above for $D=0.08$ and $r=0.1$, as depicted in
Fig. \ref{fig:Cincoh} for two frequency values:
$\Omega=0.01$ and $\Omega=0.03$ .  The picture that emerges for
$\Omega=0.03$, differs considerably from that observed for
$\Omega=0.01$. For the higher frequency case, with $t_c\approx10.4$, the
value of the solid line at $\tau=10.4$ in Fig. \ref{fig:gain:tau}~(a)
indicates that, for a considerable number of trajectories, the driving
force is not able to induce a transition during the time interval
$t_c$. Consequently, the second cumulant of $S(t)$ becomes of the order
of the distance between the attractors. On the other hand, for the lower
frequency, transitions occur for the corresponding $\tau=t_c\approx
31.4$ with almost total certainty.  This leads to a
$C_\mathrm{incoh}(0)$ for $\Omega=0.03$ much larger than for
$\Omega=0.01$.  Additionally, since still almost no transitions occur at
the intervals when the driving is absent, as shown in
Fig. \ref{fig:gain:tau} (b), nor obviously when the driving favors the
initial attractor, any unsuccessful transition is carried on until the
next period, and we would expect a long-lived
$C_\mathrm{incoh}(\tau)$. Namely, noise induced correlations persist
during a few driving periods.  These observations justify the full
behavior observed in Fig.~\ref{fig:Cincoh}.

In conclusion, we have analyzed the enhancement of SR effects in a
collective variable characterizing the response of finite sets of
interacting noisy subsystems. The cooperative effect of noise and
nonlinearity in finite sets of interacting, driven, bistable subunits
reflects in a substantial decrease in the noise level of the collective
output, $S(t)$, with respect to that observed in the output of a single
independent unit. Very large values of the SR gain can be achieved for
trains of short rectangular pulses of the type defined by
Eq.~(\ref{eq:force:def}). Even though the results presented here have
been obtained with external pulses with very brisk changes of their
amplitudes at certain instants of time, similar results can also be
observed when the pulse amplitude changes continuously, as long as there
are time intervals within a period where the amplitude changes very
slowly, followed by short time intervals where the amplitude changes
very drastically.\\

This research was supported by the Direcci\'on General de Ense\~nanza
Superior of Spain (Grant No. FIS2005-02884), the Junta de Andalucia, and
the Juan de la Cierva program of the Ministerio de Ciencia y
Tecnolog\'{\i}a (D.C.).


\end{document}